# Artificial fireball generation via an erosive discharge with tin alloy electrodes

A.L. Pirozerski[1], V.Yu. Mikhailovskii[2], E.L. Lebedeva[1], B.F. Borisov[1], A.S. Khomutova[1], I.O. Mavlonazarov[1]

[1] *Faculty of Physics, Saint-Petersburg State University*
[2] *The Interdisciplinary Resource Center for Nanotechnology of St.Petersburg State University*
*1, Ulyanovskaya str., Saint-Petersburg, 198504, Russia*
E-mail: piroz@yandex.ru

Abstract

We propose a method for generation of long-living autonomous fireball-like objects via a pulse erosive discharge between tin alloy electrodes. The objects are similar to the natural ball lightning in some properties, in particular, they have high energy density and are capable to burn through thin metal foils. The dynamics of the objects are studied using high speed videorecording. During their lifetime the fireballs generate aerogel threads. The studies of their structure by scanning electron microscopy reveal the presence of tin oxide nanoparticles and nanowhiskers.

## 1. Introduction

The problem of the physical nature of the ball lightning is one of the most interesting questions in the modern science. Although this problem was intensively investigated for a long time, both theoretically and experimentally (see [1–4] and references therein), it remains unresolved up to now.

Experiments on the ball lightning modeling carried out up to now by different authors result in the development of several methods to generate long-living autonomous luminescent objects which resemble the natural phenomenon in some properties. Conventionally, all these objects can be sorted into two groups. The first one includes dust-gas fireballs which can be produced in several ways [5–12]. At beginning of their existence these objects have a quasi-spherical shape, a size of about 10-20 cm, a noticeable surface tension and a sufficiently high luminosity that gives them visible similarity with the natural ball lightning. Plasmoids of this type can be characterized as fireballs on the base of chemically active non-ideal plasma with small dust particles of metals and/or dielectrics [11]. Their luminescence and the relative stability of the shape are due to the conversion of the chemical energy to the energy of electron or vibration excited states. The dust-gas fireballs have low energy density and, in contrast to the natural ball lightning, always undergo a relaxation and can not explode.

The fireballs of the second type (see [13–15] and references therein) are significantly smaller in size (usually, less than 1 cm in diameter) and have substantially higher energy density. An important milestone in the development of experimental techniques for creation of this type of fireballs and in understanding of their physical nature was the paper of S.E. Emelin and co-authors [13]. Several methods proposed in this paper allow more or less reproducible generation of the fireballs via an interaction of the erosive discharge plasma with metals and polymers. The fireballs demonstrated some basic features of the natural ball lightning, namely, a long time of autonomous existence and the ability to float in the atmosphere, to explode, and to burn through thin metal foils.

One type of the objects studied in [13] is so-called jumping fireballs (JF). They arose from impact of an erosional plasma, ejected from a polymethylmethacrylate capillary discharger, onto a tin anode. Their sizes were about 1-2 mm in diameter. They glowed with the bright yellow-reddish light. After falling on a laboratory table surface the fireballs jumped up and down up to 50 times with a gradual decrease of the jump height. The fireballs leaved aerogel tails which did not diffuse in the air but shrank into thin threads. Microscopy studies of the aerogel revealed a net pattern consisting of chains and agglomerates of light-gray and dark particles of sizes about 1 μm in diameter. Submicron structure of the aerogel was not studied. A qualitative model proposed [13] gives a general insight of the physical nature and properties of the jumping fireballs. However, many questions still remain open, in particular, nature of excited states, magnitude of the object specific energy, mechanisms of the metastability, the role of the polymer substance, etc.

In the present paper we propose a new method for generation of luminous objects that are very similar to the jumping fireballs described in [13]. This method does not use capillary discharger, and, therefore, polymers are not involved to the erosional plasma. The dynamics of the fireballs, features of their interaction with different targets, structure of deposits and damages on the targets are studied using high speed videorecording and optical microscopy. The aerogel structure is investigated in a wide range of scales via the optical and scanning electron microscopy.

## 2. Experiment

The discharge circuit consists of a pulse storage capacitor 1 mF x 5kV, an inductance L=2 mH integrated with a pulse ignition transformer (with the peak voltage up to 50kV) and a discharger.

The discharger included a dielectric plate with two adjustable dielectric supports carrying metallic guides for electrodes. Two piece of an industrial tin-copper alloy wire of length about 5–10 cm were used as the electrodes. The alloy composition was 97% Sn and 3% Cu. The electrodes were aligned opposite each other with a gap of 12 mm between their ends. The discharger was mounted over a laboratory table using a dielectric pillar, the distance between the electrodes and the table surface being about 30 cm.

We carried out video recording of the discharge and of the jumping fireballs using a camcoder Sony HDR-HC9 with the frame rate 50 field per second and resolution 720x576 pixels. In parallel, we use a Nikon 1 S1 digital camera for high speed video recording (640x240 at 400 fps or 320x120 at 1200 fps).

The optical microscopy studies were performed with use of a Micmed 2000R digital USB microscope which can acquire 1600x1200 pixel images with maximal magnification $200^x$.

The scanning electron microscopy studies were carried out in the Interdisciplinary Resource Center for Nanotechnology of St.Petersburg State University using field-emission SEM Zeiss Merlin. Accelerating voltage was 8 kV and in-lens detector of secondary electrons was used.

## 3. Experimental results and discussion.

**3.1. Overview of a typical experiment**. Frames from a videorecord of a typical experiment on the jumping fireballs generation are shown in Figs. 1–3 below. The initial voltage of the capacitive storage was 2.9 kV. The discharge duration was about 25 ms with the final voltage of the storage not exceeding 50 V. At ~20 ms after the current breaking a luminous plasmoid and a multitude of fast moving bright spark-like objects are visible (Fig. 1a). After ~50 ms the plasmoid fade away and turns into a dust cloud which is still visible hundreds of milliseconds after the current breaking (the left-top area on Fig. 1b). The spark-like objects fall on the laboratory table, some of them split at collisions with the table surface (Fig. 1b). The objects leave bluish-gray thread-like tails consisting of the tin oxide aerogel which are floating up in the atmosphere (Fig. 1b). The tails undergo a gradual turbulization and breaks up into smaller pieces and shrink.

After falling on the table, the objects move on the surface either gliding or jumping to height up to few centimeters. The combined image of a sequence of 240 frames (total duration 0.6 s) of the high speed video, shown in Fig.2, visualizes trajectories of several jumping fireballs. It should be noted that JFs move actively, i.e. they have a source of the internal energy which transforms to the mechanical one. In particular, the height of the jumps can increase, some of the gliding objects can start jumping, and "gliding" and "jumping" parts of the trajectory of the JFs can alternate. Note also that at collision with an object a jumping fireball can break up into a few new jumping fireballs (usually 2-3).

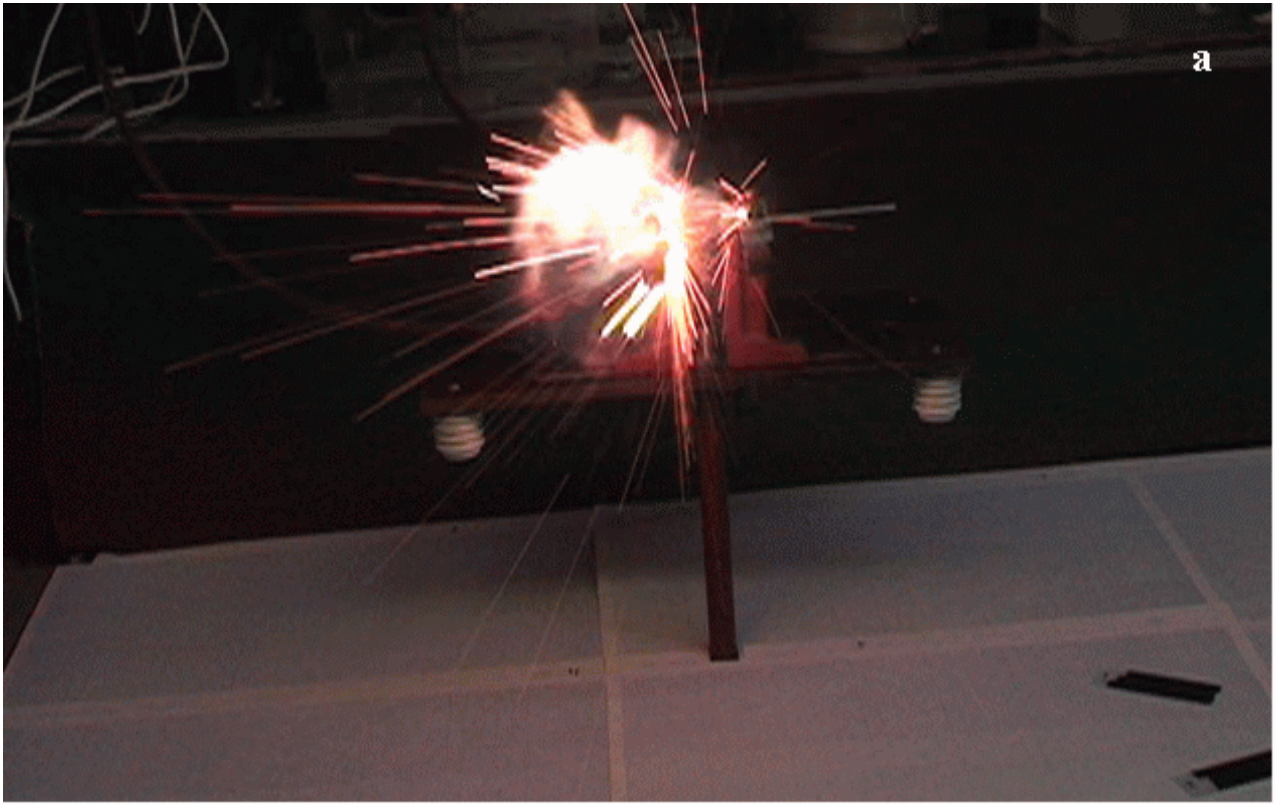

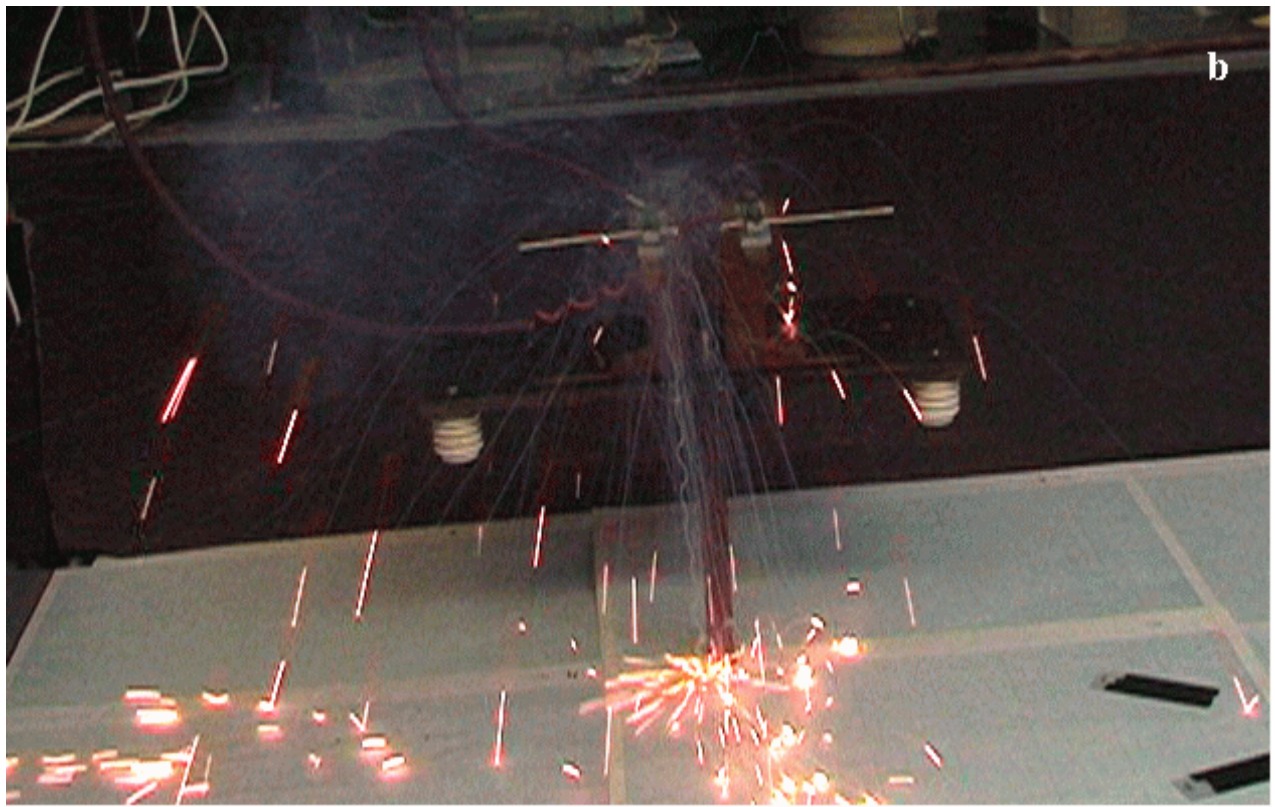

Fig.1. Frames from a videorecord of an experiment on JF generation taken at ~20 ms (a) and at ~280 ms (b) after the discharge current breaking. For the right (b) image the exposure was adjusted to visualize aerogel tails of the objects.

A close view of JFs is presented in Fig.3. The objects have a comet-like shape with quasi-spherical bright kernel and a pronounced tail resembling a smoke. But unlike the smoke the tail does not diffuse in the air, on the contrary, it shrinks and breaks into relatively short thin threads that can float in the atmosphere during several minutes before they finally deposit on the surrounding objects.

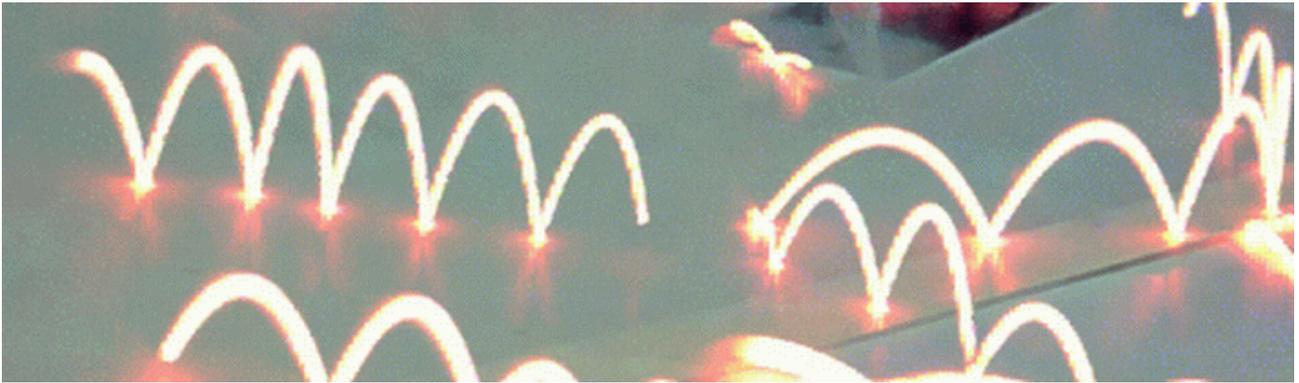

Fig.2. The combined image of a sequence of 240 frames from the high speed videorecord of the experiment. The beginning of the sequence corresponds to 1012 ms after the current breaking, and the sequence duration is 0.6 s.

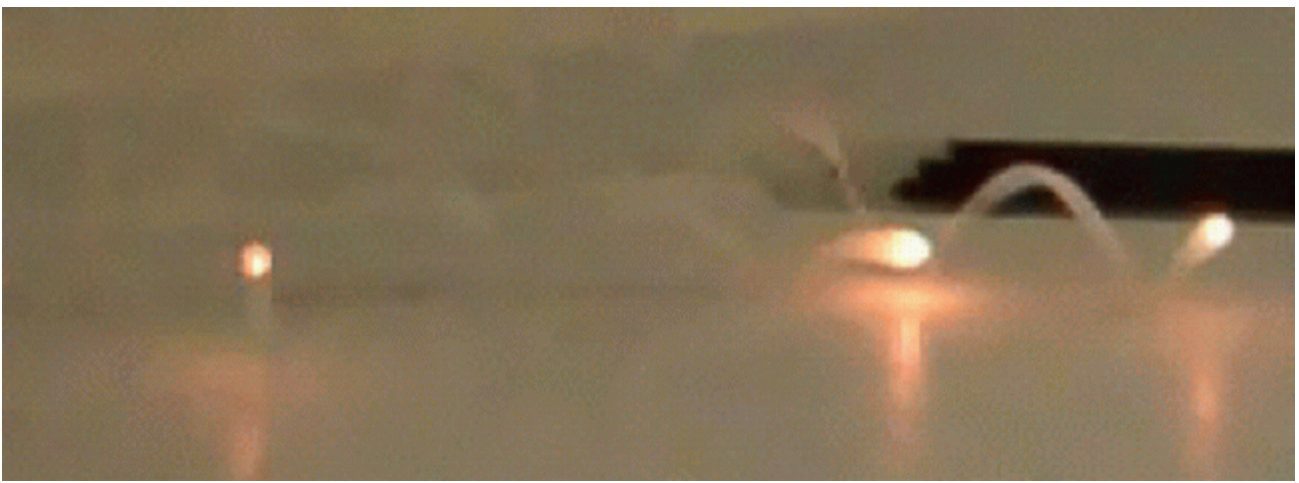

Fig.3. A close view of jumping fireballs (1080 ms after the current breaking).

3.2. **Impact on targets.** At the contact with an object the jumping fireballs produce traces and/or damages the character of which depends on material constituting the object. Typical examples of the traces are presented in Fig.4.

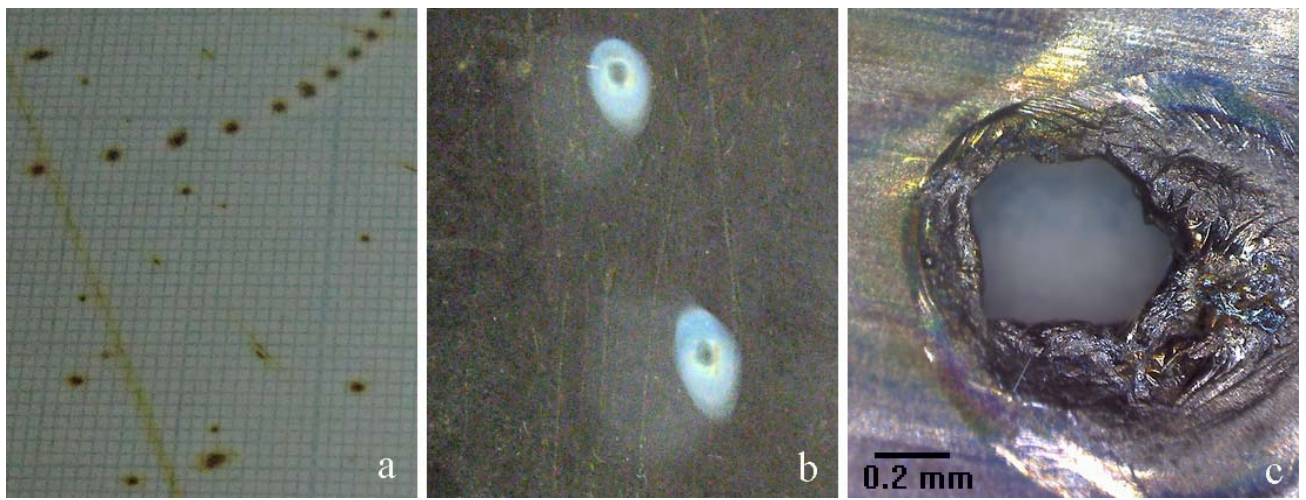

Fig.4. Traces and damages produced by the JFs on different materials: (a) – on a coordinate paper (the small division is 1 mm); (b) – on the surface of a getinaks sheet; (c) – on an aluminium foil of thickness 9 μm.

Fig. 4a shows two types of traces produced by jumping fireballs on a coordinate paper. The first one that can be seen on the top of the image consists of separate brown spots placed at points of the contact of a jumping fireball with the paper. The spots are darker in center and look like scorches resulting from the contact with a very hot object.

The traces of the second type that look like brownish lines are produced by the JFs when they glide on the paper surface. An example of such a trace is shown on the left part of Fig. 4a.

On the surface of the getinaks or the textolite JFs produce traces with more complex structure (Fig.4b). In the inner part of the trace there is an oval brown sport edged with a thin white ring. The ring is surrounded with a bluish-grey area of the oval form. From its outside, in the direction, which usually is opposite to the projection of the jumping fireball velocity on the surface, there is another small region of a grey color and a crescent form. The described system of domains will be referred to as the main part of the trace. Its greatest dimension usually does not exceed 4–5 mm. Additionally, a translucent oval or oblong veil up to 1 см in size is often present. The main part of the trace is usually situated near an edge of the veil. It should be noted that getinaks or textolite surfaces can be easily cleaned of the traces using a cloth.

At collision with a thin metal foil the jumping fireballs are capable to burn it through. Fig.4c shows a piece of 9 μm thick aluminium foil that was damaged by a JF. The size of the hole is about 0.5 mm but sometimes it can amount to 2-3 mm. Near the hole edges the metal was deformed and somewhere reflowed by heating. At longer distance from the hole the surface of the foil is tinted to different colors followed by a grayish domain (not visible in Fig.4c) resembling the traces shown in Fig.4b.

**3.3. Structures of jumping fireball remains and aerogel threads.** Fig.5a shows microscope image of a jumping fireball residue. It can be seen that the residue consists of two parts bound together. As it was already noted, at a collision with an object a JF can split up into several parts. In this case, probably, the jumping fireball had lost activity before the splitting was completed. On the surface of the smaller part of the JF a crannied crust is present which consists mainly of the tin oxide. There are small pieces of the aerogel in several spots on the crust surface. On the larger part of the JF a hexagonal cavity is present which probably results from a flow of melted tin to the smaller part.

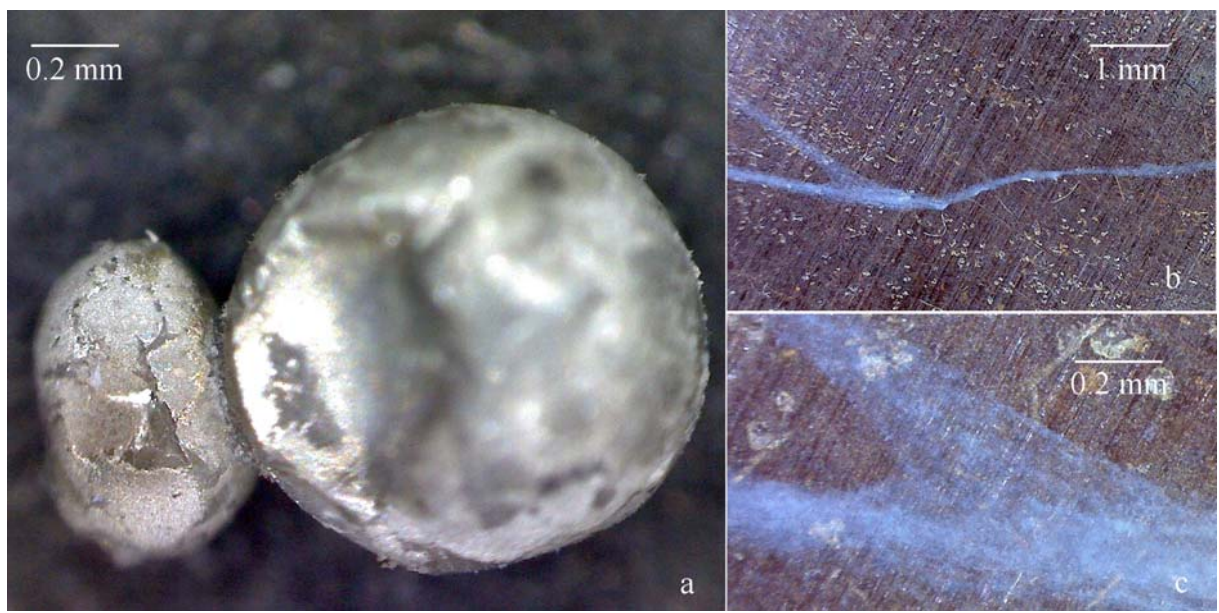

Fig.5. (a) Residue of a jumping fireball. (b, c) Microscope images of a piece of an aerogel thread on the surface of a textolite plate at different magnifications. Scales are indicated on the images.

Fig.5 (b, c) show images of an aerogel thread on the surface of a textolite plate obtained using the optic microscope. Out of the branching area the thread diameter is about 0.1 mm. As can be seen from Fig.5c, the thread has a non-homogeneous, loose structure and is semi-transparent.

Detailed studies of the aerogel structure on the micron and submicron scales were performed using scanning electron microscopy. SEM-images of an aerogel sample at different magnifications from $1500^x$ to $151000^x$ are shown in Fig.6. As may be seen from Fig.6a, on scales of about tens microns the aerogel has a stochastically non-homogeneous cellular fibrous structure with a high dispersion of the mesh sizes. At higher magnification (Fig.6b) individual nanoparticles became discernible but the general pattern of the substance distribution is very similar to the one shown in Fig.6a. This allows considering the aerogel substance on the corresponding scale range as a (stochastic) fractal. The nanoparticles can either be interconnected with each other by nanowhiskers or form agglomerates (Fig. 6c). The characteristic size of the nanoparticles is 35-90 nm. Some of the nanoparticles have clearly visible facets and edges (Fig. 6d) which implies the crystalline structure. Note that such shapes of the nanoparticles are typical for $SnO_2$, while SnO forms lamellate nanocrystals [16].

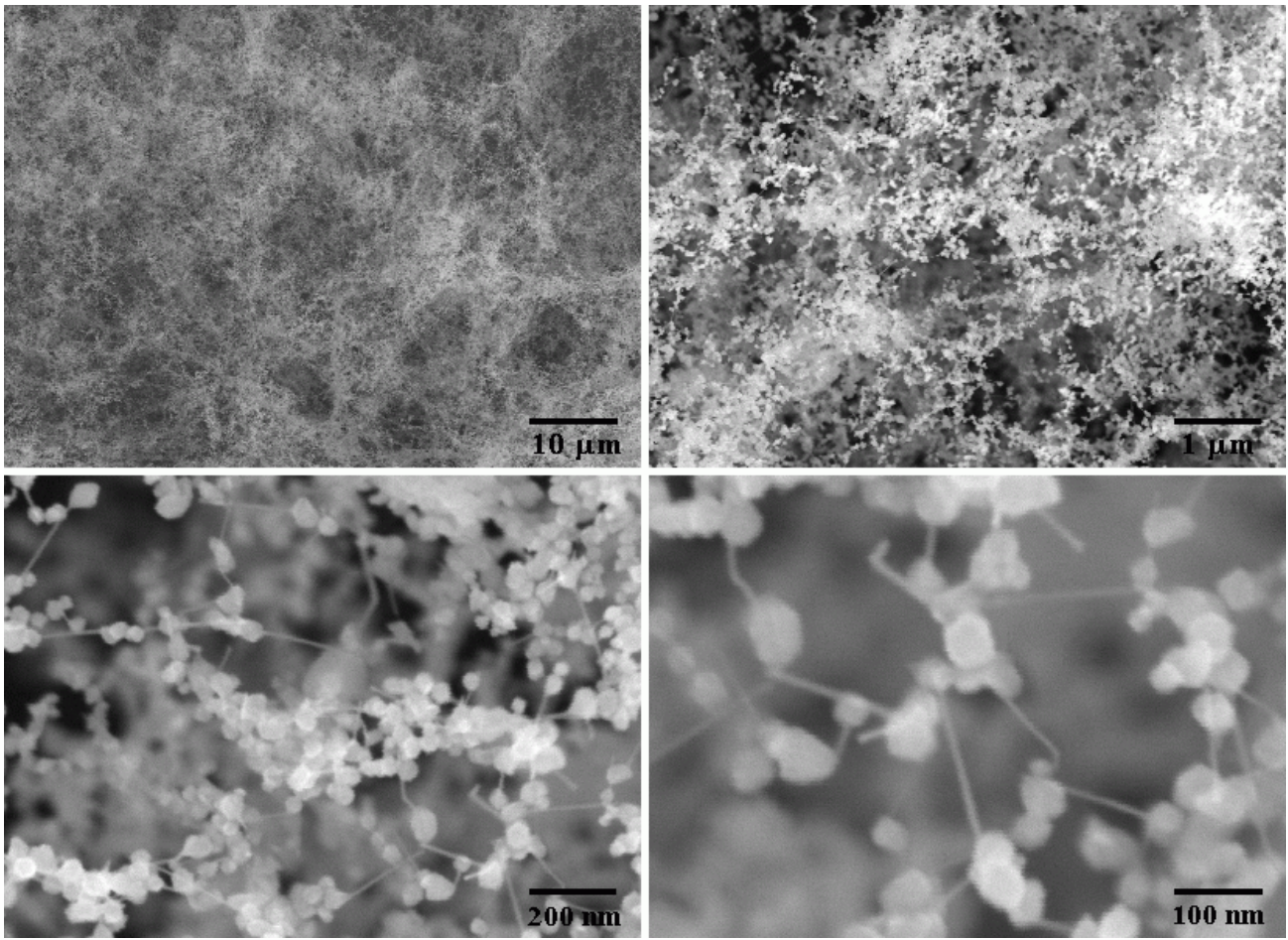

Fig. 6. SEM images of the aerogel substance. [IRC for Nanotechnology of St.Petersburg State University]

The typical diameters of the nanowhiskers are 10-20 nm while their lengths amount to 200 nm and more. Some of the nanowhiskers have angle-like flexures and even have shape of a polygonal line with angles between the segments attaining 90 degrees (see the top-center area in Fig. 6d). (It should be noted that formation of nanowhiskers by tin evaporation in a low-pressure argon-oxygen mixture was reported in [17, 18].)

**3.4. Qualitative explanation of the active dynamics of the jumping fireballs.**

As it was already noted above the dynamics of the jumping fireballs is active which means that the mechanical energy of the fireballs can increase. Up to now detailed explanation of this phenomenon is lacking. Nevertheless, main underlying processes can be qualitatively described in the following way.

After the interelectrode gap breakdown and the onset of the non-stationary arc discharge a heating and an intense erosion of the electrodes take place. The electrode material in the form of a vapor with little drops of the liquid tin enters into the plasma where a not understood process of "charging" and activation of the eroded substance by the chemically active plasma occurs [13]. Although further studies are required to understand the physical nature and features of this process, so far we suppose that the process leads (in particular) to the conversion of the input electric energy into some other forms, probably into collective excitations of the electron subsystem, into the energy of multicharged ions or free radicals, etc.

The "charged" and activated tin drops undergo non-equilibrium oxidation resulting in formation of tin oxide $SnO_2$ (perhaps, with intermediate formation of SnO and its subsequent oxidation). Since the melting temperature of $SnO_2$ is ~1903 K, whereas one of the tin is only ~505 K, the tin oxide is capable not only to form the aerogel tail of the jumping fireball, but also to precipitate in the solid state on the surface of the liquid tin drop. This results in the formation of a layer which impedes the oxygen inflow to the surface of the jumping fireball. A collision of the jumping fireball with the surface of the laboratory table results in the mechanical destruction of a part of the layer. This, in its turn, leads to an abrupt increase of the oxidation rate and to an intense formation of the aerogel which is ejected from the tin drop. Corresponding reactive force and an overpressure near the surface give rise to the increase of the jumping fireball mechanical energy.

**Conclusion**. In the present paper we propose the method for generation of autonomous long living luminous objects – jumping fireballs – via an erosive discharge between tin alloy electrodes. The jumping fireballs have a partial similarity with the natural ball lightning, in particular, they have sufficiently bright luminescence, can move actively and burn through thin metal foils. During their lifetime the jumping fireballs generate tin oxide aerogel tails that, unlike to a smoke, do not diffuse but shrink to thin threads floating in air. The studies of the aerogel thread structure using SEM reveal the presence of tin oxide nanoparticles with irregular polyhedron forms and size of 35–90 nm and nanowhiskers (~10-20 nm in diameter). A qualitative explanation of the active dynamics of the jumping fireballs is given.

**Acknowledgments**

This work was partially supported by Saint-Petersburg State University, grant 11.0.62.2010. The SEM studies were carried out in the Interdisciplinary Resource Center for Nanotechnology of St.Petersburg State University.

**References**

[1] Barry, J. D., 1980, "Ball Lightning and Bead Lightning", Plenum Press, New York and London.
[2] Stakhanov, I. P., 1985, "The Physical Nature of Ball Lightning", Energoatomizdat, Moscow (in Russian).
[3] Smirnov, B.M., 1993, "Physics of Ball Lightning", Physics Reports, 224(4), pp. 151–236.
[4] Ball Lightning in the Laboratory, Khimiya, Moscow, 1994 (in Russian).
[5] Powel, J. R., Finkelstein, D., 1970, "Ball Lightning", Amer. Scientist, 58(3), pp.262–280.
[6] Furov, L.V., 2005, "Generation of Autonomous Long-Lived Plasma Objects in Free Atmosphere", Technical Physics, 50(3), pp. 380–383.
[7] Egorov, A. I., Stepanov, S. I., Shabanov G. D., 2004, "Laboratory demonstration of ball lightning", Physics-Uspekhi, 47(1), pp. 99–101.
[8] Emelin, S.E., Pirozerski, A.L., Egorov, A. I., Stepanov, S. I., Shabanov, G. D., Bychkov, V.L., 2002, "Ball lightning modeling via electrical discharge through the surface of weak water solution", Proc. 9-th Russian Conf on Cold Nuclear Transmutation and Ball Lightning, Moscow, pp. 240–248.

[9] Pirozerski, A.L. and Emelin, S.E., 2002, "Long-living plasma formations arising from metal wires burning", preprint arXiv:physics/0208062.
[10] Emelin, S.E., Pirozerski, A.L., 2006, "Some questions on the dense energy plasma-chemical fireball", Chemical Physics, 25(3), pp.82–88.
[11] Emelin, S.E., Pirozerski, A.L., Vassiliev, N.N., 2006, "Dust-gas fireball as special form of electric erosive discharge afterglow", Proceedings of Ninth International Symposium on Ball Lightning (ISBL-06), G. C.Dijkhuis, ed.; Preprint: http://arxiv.org/abs/physics/0604115.
[12] Pirozerski, A.L., and Emelin, S.E., 2006, "Long-living plasmoids generation by high-voltage discharge through thin conducting layers", Proceedings of Ninth International Symposium on Ball Lightning (ISBL-06), G. C.Dijkhuis, ed.; Preprint: http://arxiv.org/abs/physics/0606260.
[13] Emelin S.E., Semenov V.S., Bychkov V.L., Belisheva N.K., Kovshik A.P., 1997, "Some objects formed in the interaction of electrical discharges with metals and polymers", Tech. Phys, 42(3), pp. 269–277.
[14] Bychkov A.V., Bychkov V.L., Timofeev I.B., 2004, "Experimental production of long-lived luminous organic polymeric objects", Technical Physics, 49(1), pp.128–132.
[15] Emelin, S.E., Bychkov, V.L., Astafiev, A.M., Kovshik, A.P., Pirozersky, A.L., 2012, "Plasma Combustion Nature of Artificial Ball Lightning", IEEE Transactions on Plasma Science, 40(12), pp. 3162–3165.
[16] Kaito, C. and Shiojiri, M., 1982, "Structure and Coalescence Growth of Smoke Particles of Tin Oxides and Lead Oxides Prepared by Burning the Metals in Mixture of Ar and $O_2$", Jap. J. Appl. Phys., 21(10), pp.1404–1408.
[17] Kim, H. W., Lee, J. W., Shim, S.H. and Lee, C., 2007, "Controlled Growth of $SnO_2$ Nanorods by Thermal Evaporation of Sn Powders", J. Korean Phys. Soc., 51(1), pp.198–203.
[18] Kim, H. W., Shim, S.H., 2007, "Temperature-controlled fabrication of $SnO_2$ nanoparticles via thermal heating of Sn powders", Appl. Phys. A, 88, pp. 769–773.